# Quantum Amplitude Estimation for Travel Time Estimation in Stochastic Vehicle Routing Problems

Xingyue Wang and Monika Filipovska

***Abstract*—Solving the Vehicle Routing Problem (VRP) in Stochastic Transportation Networks (STNs), a core task in Intelligent Transportation Systems (ITS), introduces estimation challenges for stochastic path travel times and the resulting VRP objective function. These challenges have typically been addressed through computationally expensive sampling-based techniques such as Monte Carlo simulation, whose performance depends on sample size, the sampling strategy, and the underlying travel time distributions. To address these issues, this study proposes and validates a quantum computing technique, Quantum Amplitude Estimation (QAE), for path-level travel time estimation in STNs. Without relying on sampling or prior assumptions of the travel time distribution, the proposed framework encodes all feasible travel time realizations into a quantum superposition, enabling a theoretical quadratic speed-up over Monte Carlo simulation. Four QAE variants are implemented in IBM's Qiskit framework, namely Canonical AE (CAE), Iterative AE (IAE), Maximum Likelihood AE (MLAE), and Faster AE (FAE), together with four rotation-angle scaling strategies for handling different discrete travel time distributions. Experiments on a small-scale STN show that the choice of scaling method and rotation-angle range significantly affects estimation accuracy, while the four QAE variants produce comparable estimates across all tested conditions, with IAE exhibiting the most stable overall performance. The results provide practical guidance on parameter selection for future hybrid quantum-classical optimization frameworks in ITS applications.**

## I. Introduction

The classical Vehicle Routing Problem (VRP) is a fundamental NP-hard combinatorial optimization problem in transportation networks [1], which seeks optimal routes for a fleet of vehicles under given operational parameters, particularly travel times. In intelligent transportation systems (ITS), VRP plays an important role in real-time logistics operations and transportation decision-making, such as fleet management, last-mile delivery, and emergency response management [2]. However, real-world ITS environments are frequently disrupted by traffic congestion, incidents, and adverse weather, introducing stochastic travel time variations and network dynamics that the deterministic VRP cannot adequately capture [3]. This has motivated increasing research on the VRP with Stochastic Travel Times (VRPSTT) [4].

Incorporating uncertainty, however, complicates the VRP formulation, as the resulting objective functions are analytically intractable and computationally costly to evaluate. Classical studies commonly address this issue through Monte Carlo simulation within simulation-based optimization (SBO) [5], [6]. However, the computational burden of repeatedly estimating path or route costs through such methods grows rapidly with network scale, routing complexity, and desired estimation accuracy, becoming a critical bottleneck for real-time ITS operations and dynamic logistics services.

Quantum computing has recently emerged as a powerful paradigm to address such computational bottlenecks, applying quantum parallelism to efficiently model complex probability distributions and stochastic processes [7]. Despite this potential, its application to advanced transportation problems remains largely unexplored, creating a two-fold research gap. On one hand is the undiscovered potential of quantum algorithms to provide new solutions to numerous transportation problems by taking advantage of unique quantum properties, such as superposition and interference [8]. On the other hand, current quantum computers are still in an early stage of development, known as Noisy Intermediate Scale Quantum (NISQ) systems [9] and a key open challenge for ITS research is to identify transportation problems for which quantum methods can provide verifiable improvements over existing classical approaches.

From an ITS perspective, efficient estimation of stochastic travel times is essential to uncertainty-aware routing and logistics planning [6], [10]. To address this gap, this study proposes a method based on Quantum Amplitude Estimation (QAE) [11] for path-level travel time estimation in VRPSTT. By encoding all possible realizations of a path's travel time into a quantum superposition, the proposed QAE-based framework eliminates the dependence on classical sampling and provides a theoretical quadratic speed-up over Monte Carlo methods [12]. Furthermore, the estimator operates without prior assumptions on the travel time distributions, significantly improving the generality and applicability to diverse ITS scenarios.

As the first application of QAE for expected path travel time estimation in Stochastic Transportation Networks (STNs), this work evaluates four QAE variants using IBM's Qiskit platform [13]: Canonical [11], iterative [14], maximum likelihood [15], and faster [16] amplitude estimation (AE). Through experiments on a small-scale STN, we analyze and compare the computational accuracy and error performance of these four variants.

This work was supported by the National Science Foundation under Award No. 2520130. *(Corresponding author: Monika Filipovska)*

Xingyue Wang and Monika Filipovska are with the School of Civil and Environmental Engineering, University of Connecticut, Storrs, CT 06269 USA (e-mail: xingyue.wang@uconn.edu; monika.filipovska@uconn.edu).

It is important to note that this work focuses on evaluating the feasibility and performance of QAE for the core expectation-estimation subproblem in stochastic routing problems, rather than developing a complete VRP solver. Since SBO remains the dominant approach in this domain, accelerating expectation evaluation constitutes a natural entry point for quantum-enhanced methods.

The main contributions of this paper are: (1) A QAE-based path travel time estimator is formulated for STNs without parametric assumptions on travel time distributions. (2) Four QAE variants are implemented and compared in terms of estimation accuracy and error characteristics. (3) Encoding and different scaling strategies are evaluated to reveal their effects on estimation performance. (4) Simulation results demonstrate the feasibility of QAE for accurate path travel time estimation and show its potential for ITS applications. Collectively, this work extends quantum algorithm applications to STNs and provides a foundation for future hybrid quantum-classical methods for stochastic VRPs.

The remainder of this paper is organized as follows. Section II reviews related literature. Section III details the QAE-based path travel time estimation framework. Section IV presents experimental setup and design. Section V reports and analyzes results, and Section VI concludes the paper.

## II. Literature Review

### *A. Classical Approaches for VRPSTT*

To capture travel time uncertainty in VRP, conventional literature predominantly relies on simulation-based strategies where Monte Carlo sampling estimates expected routing costs. These models typically assume that travel time follows a specific parametric distribution, such as lognormal [17], normal [18], or Gamma distributions [19], with analytical alternatives such as the Wilkinson method under similar assumptions [20]. Building on this foundation, sampling-based strategies have been widely applied across VRP variants, including the Probabilistic Time-Dependent VRP [21] and adaptive sampling for the stochastic VRP with time-window [22].

Overall, prevailing classical methods for VRPSTT face two main limitations. First, they generally rely on prior parametric assumptions on travel time distributions to derive path-level characteristics. Second, their dependence on Monte Carlo simulation places most VRPSTT formulations within the SBO category. The estimation precision remains highly sensitive to sample size, requiring an inefficiently large number of simulation replications to support real-time ITS routing decisions.

### *B. Quantum Computing for Stochastic Modeling and Optimization*

Built upon one of the most influential quantum algorithms, Grover's search [23], Quantum Amplitude Estimation (QAE) offers a solid mathematical framework for accelerating estimation of unknown distribution values with bounded variance [24]. It has achieved validated success in simulating stochastic processes [25] and evaluating financial options [26]. Notably, QAE demonstrates substantial advantages in discrete stochastic processes (DSPs) [27] and SBO, where it provides a quadratic speed-up over Monte Carlo simulation. Specifically, QAE requires $O(1/\epsilon)$ quantum queries to achieve an estimation precision that would necessitate $O(1/\epsilon^2)$ classical samples for a target error $\epsilon$ [12].

In the context of transportation problems, existing literature has focused on the combinatorial optimization of VRPs using the Quantum Approximate Optimization Algorithm (QAOA) [28], Variational Quantum Eigensolver [29], or quantum annealing via Quadratic Unconstrained Binary Optimization (QUBO) [30]. However, these studies focus exclusively on combinatorial path optimization in deterministic VRPs, leaving open the question of how quantum methods can address path travel time estimation in STNs.

Although a direct quantum method for VRPSTT is still lacking, foundational work has begun to address the representation of stochasticity itself. For the stochastic VRP, Filipovska [31] proposed encoding stochastic travel times as quantum states and simulating their probability distributions through quantum circuits, providing an alternative to large-scale sampling for characterizing travel time distributions. Building on this, this study develops a QAE-based framework for path travel time estimation in VRPSTT and demonstrates both its theoretical basis and practical effectiveness.

## III. Methodology

This section details the proposed QAE-based framework for estimating expected path travel times in STNs. It first introduces the quantum principles and analogy between QAE and classical Monte Carlo methods, then formalizes the STN and the estimation problem. Finally, it presents the QAE-based methodology, four QAE variants, four rotation-angle scaling methods for travel time mapping, and the scalability and hardware limitations under current NISQ devices.

### *A. QAE Primer for ITS Researchers*

QAE estimates the expected value of a random variable without generating repeated independent samples. Instead, it utilizes quantum parallelism to encode all possible realizations into a single quantum state. This subsection introduces the fundamental quantum concepts behind QAE to ITS researchers through analogies with classical Monte Carlo concepts, clarifying how stochastic modeling tasks are mapped onto the quantum framework.

The classical Monte Carlo method approximates the expected output $\mu$ of a randomized algorithm $\mathcal{M}$ by generating $N$ independent samples and computing their average as the estimator $\tilde{\mu}$ [32]. The estimator converges asymptotically as $N$ increases, necessitating extensive sample sizes to achieve high precision. This classical approach serves as a baseline to demonstrate how QAE eliminates both repeated sampling and parametric distribution assumptions.

Unlike a classical bit that is restricted to binary 0 or 1, the fundamental unit of quantum information is the qubit,

represented by the basis state $|0\rangle$ and $|1\rangle$. A qubit can exist in a linear combination of these basis states, written as $|\psi\rangle = \alpha|0\rangle + \beta|1\rangle$, where the complex coefficients $\alpha$ and $\beta$ represent probability amplitudes satisfying $|\alpha|^2 + |\beta|^2 = 1$ [33]. This property, known as superposition, allows the qubit to exist in both basis states simultaneously [33]. The squared magnitudes $|\alpha|^2$ and $|\beta|^2$ give the probability of collapsing into the states $|0\rangle$ and $|1\rangle$, rerspectively, upon measurement.

While Monte Carlo simulation enumerates traffic scenarios one at a time, superposition encodes the entire probability space of feasible scenarios within a single quantum state, enabling parallel rather than sequential processing. In the proposed QAE framework, the quantity of interest, which in this study is the expected path travel time, is encoded directly into the amplitude of a specific component of the quantum state. This differs from Monte Carlo estimation, where the expectation is derived by averaging discrete sample outcomes.

Quantum state transformations are implemented through quantum circuits, which apply structured sequences of unitary gates to manipulate qubits [33]. Each gate performs a predefined linear operation, enabling the controlled preparation, evolution, and manipulation of the quantum states required by algorithms such as QAE. Measurement subsequently collapses a superposition state $|\psi\rangle = \sum_k \alpha_k |k\rangle$ into a distinct basis state $|k\rangle$ with a probability of $|\alpha_k|^2$. To resolve the final amplitude with high confidence, the algorithm executes the circuit for multiple repetitions or shots [11]. Unlike classical Monte Carlo where repeated sampling is the primary computational driver, QAE maintains the superposition throughout the circuit and defers measurement to the final stage to extract the computed amplitude.

The primary difference between QAE and Monte Carlo simulation lies in the mechanism used to estimate expectation values. In the context of VRPSTT, Table I provides an interpretation of how the stochastic travel time estimation problem is represented withing the QAE framework.

TABLE I
CONCEPTUAL MAPPING BETWEEN TRANSPORTATION AND QUANTUM CONCEPTS.

| Transportation Concept | Quantum Equivalent |
|---|---|
| Travel time realization | Computational basis state |
| Probability of realization | Squared amplitude of the corresponding basis state |
| Path cost function | Observable |
| Expected path travel time | Rescaled QAE estimate |
| Monte Carlo samples | Quantum queries |

### *B. Stochastic Transportation Network Representation*

STNs exhibit complex spatial and temporal dynamics, typically modeled as graphs in which each road segment carries a link travel time distribution that captures its probabilistic nature. Integrating QAE into such networks first requires a formal definition of network uncertainty. This section therefore defines the random variables, formulates path-level travel times in a discrete representation, and then incorporates them into the VRP formulation.

Let an STN be represented by a directed graph $G = (V, A)$, where $V = \{0, \dots, v\}$ is a set of nodes and $A = \{(p, q), p \in V, q \in V, p \neq q\}$ is a set of directed links, with $(p, q)$ denoting a link from node $p$ to node $q$. Network stochasticity is defined over a probability space $(\Omega, \mathcal{F}, P)$, where $\Omega$ is the sample space of all network realizations, $\mathcal{F}$ is the set of events, and $P$ is the associated probability measure.

The travel time of each link $(p, q) \in A$ is modeled as a discrete random variable $T_{pq}$. For a given path $h$ consisting of $n$ directed links, the links are indexed according to their order along the path as $l = 1, \dots, n$. The $l$-th link is denoted by $(p_l, q_l)$, and its travel time is written as $T_l \coloneqq T_{p_l q_l}: \Omega_l \to \mathbb{R}^+$, where $\Omega_l$ is the sample space of link $l$. The path travel time $T_h$ is the sum of the link travel times along $h$: $T_h = \sum_{l=1}^{n} T_l$. Let $\Omega_h = \Omega_1 \times \dots \times \Omega_n$ denote the joint realization space of path $h$, and let $\omega \in \Omega_h$ be a specific joint realization. The expected path travel time can be expressed as:

$$\mathbb{E}[T_h] = \sum_{\omega \in \Omega_h} t(\omega) P(\omega) \quad (1)$$

where $t(\omega)$ is the path travel time under realization $\omega$ and $P(\omega)$ is its probability.

Following Filipovska's foundational work [31] and adapting the DSP formulation of Blank et al. [27] to STNs, the discrete random travel time $T_l$ of the $l$-th link takes at most $k$ realizations $t_{l,0}, \dots, t_{l,k-1} \in \mathbb{R}^+$, so that $T_l(\Omega_l) = \{t_{l,0}, \dots, t_{l,k-1}\}$. With the index set $K = \{0,1, \dots, k-1\}$, each joint realization $\omega$ is uniquely identified by an index vector $j = (j_1, \dots, j_n)^T \in K^n$, where $j_l$ selects the realized value of link $l$. The corresponding realization vector is $x(j) = \left(t_{1,j_1}, \dots, t_{n,j_n}\right)^T$ [27], with the $l$-th component $x_l(j) = t_{l,j_l}$. Collecting the link travel times into $T = (T_1, \dots, T_n)^T$ and assuming mutual independence, the joint probability factorizes as $P[T = x(j)] = \prod_{l=1}^{n} P(T_l = t_{l,j_l})$ with $\sum_{j \in K^n} P[T = x(j)] = 1$. Let $f(\cdot)$ denote the path cost function, (1) can be written in index form as:

$$\mathbb{E}[T_h] = \sum_{j \in K^n} f\left(\sum\nolimits_{l=1}^{n} x_l(j)\right) P[T = x(j)] \quad (2)$$

In this work, the estimation is intended to be embedded in a standard VRPSTT formulation. Rather than solving the global combinatorial routing directly, this work targets the primary computational bottleneck: evaluating the expected travel time of candidate paths. The resulting estimates can then serve as reliable inputs for subsequent classical or quantum optimization algorithms, thereby supporting downstream solution of the VRPSTT.

Let $x_{pq}$ denote the binary decision routing variable that equals 1 if link $(p, q)$ is traversed and 0 otherwise. Using the link travel times $T_{pq}$ defined above, the VRPSTT objective function is formulated as $min\, \mathbb{E}\left[\sum_{(p,q) \in A} T_{pq} x_{pq}\right]$ or $min\, \mathbb{E}\left[\sum_{c \in C} \sum_{(p,q) \in A} T_{pq} x_{pq}^c\right]$, where $C$ is a set of vehicles, and $x_{pq}^c \in \{0,1\}$ equals 1 if vehicle $c$ traverses directed link $(p, q)$, and 0 otherwise.

This formulation is subject to standard routing constraints, ensuring that each node is visited exactly once, depot return policies are enforced, and subtours are eliminated, while specific capacity limits are omitted to isolate the estimation performance. Additional constraints such as vehicle capacity or customer demand are not considered in this study.

### *C. Theoretical Framework of QAE*

Formally, QAE is designed to estimate the amplitude of a specific quantum state. Let $\mathcal{H}$ denote the Hilbert space of a quantum system. A Boolean function $\mathcal{X}: Z \to \{0,1\}$ partitions $\mathcal{H}$ into a "good" subspace, spanned by the basis states $|x\rangle \in \mathcal{H}$ for with $\mathcal{X}(x) = 1$, and its orthogonal complement, the "bad" subspace [11]. Any state $|\Psi_f\rangle$ in $\mathcal{H}$ can then be written as $|\Psi_f\rangle = |\Psi_0\rangle + |\Psi_1\rangle$, where $|\Psi_0\rangle$ and $|\Psi_1\rangle$ the projections of $|\Psi_f\rangle$ onto the bad and good subspaces, respectively. Consider a unitary operator $\mathcal{A}$ acting on $d$ qubits that prepares the state $\mathcal{A}|0\rangle^{\otimes d} = \sqrt{1-a}|\Psi_0\rangle + \sqrt{a}|\Psi_1\rangle$, where $a \in [0,1]$ is the probability of measuring the good subspace $|\Psi_1\rangle$. The goal of QAE is to produce an estimate $\tilde{a}$ of the target quantum amplitude $a$ [11].

As the original formulation of QAE, CAE [11] combines Quantum Phase Estimation (QPE) [34] with amplitude amplification. The algorithm uses two quantum registers: an evaluation register with $m$ qubits in equal superposition, and a state register with $d$ qubits prepared by $\mathcal{A}$, giving a total of $M = 2^m$ quantum samples. This step encodes the amplitude information into the phases of the evaluation qubits. The Grover operator is then applied $M-1$ times in a controlled manner to amplify the amplitude, and it is defined as: $\mathcal{Q} = \mathcal{A}\mathcal{S}_0\mathcal{A}^\dagger\mathcal{S}_{\Psi_0}$, with $\mathcal{S}_0 = \mathbb{I}_d - 2|0\rangle^{\otimes d}\langle 0|^{\otimes d}$ and $\mathcal{S}_{\Psi_0} = \mathbb{I}_d - 2|\Psi_0\rangle\langle\Psi_0|$ as reflection operators. Performing an inverse Quantum Fourier Transform (QFT) [35] on the evaluation register yields a measured integer $z \in \{0, \ldots, M-1\}$, from which the final amplitude estimate is computed as $\tilde{a} = \sin^2(z\pi/M)$. This estimate satisfies $|a - \tilde{a}| \le \pi/M$ with probability at least $8/\pi^2$ [12]. Fig. 1 illustrates the above steps of CAE using a four-qubit circuit with three evaluation qubits (i.e., $m = 3$) and one state qubit (i.e., $d = 1$).

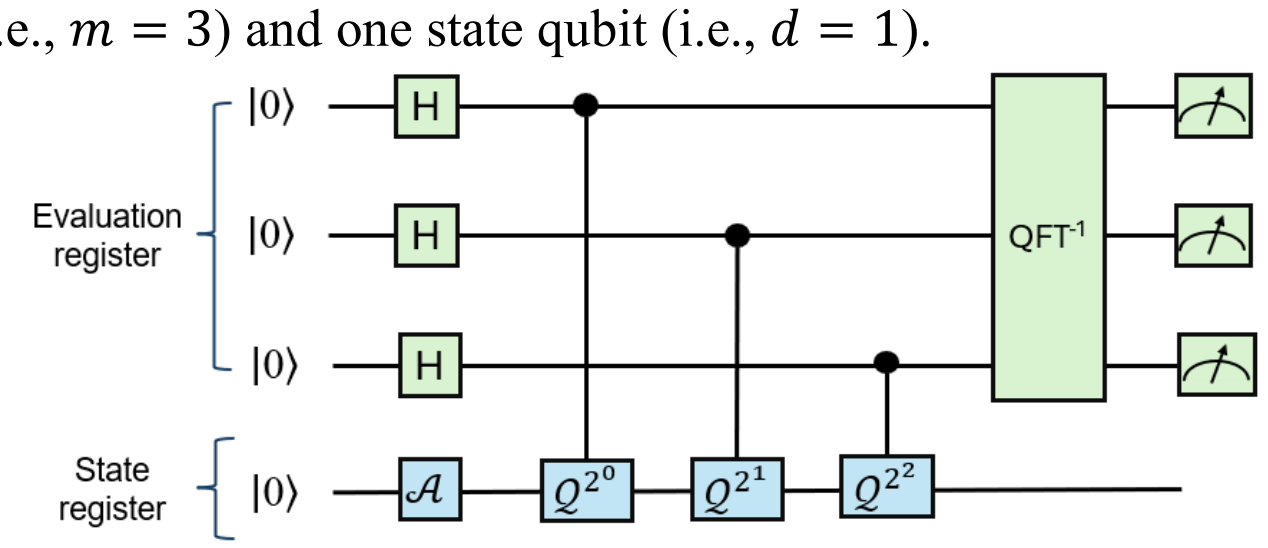

**Fig. 1.** A four-qubit toy CAE circuit with a three-qubit evaluation register and a one-qubit state register.

Within the context of this study, the QAE workflow can be summarized in four steps: (1) A quantum circuit prepares a superposition state encoding all feasible path travel times for a given VRPSTT case. The operator $\mathcal{A}$ maps the initial state to a superposition in which each basis state corresponds to a specific path realization, and the squared amplitude gives its probability. (2) The operator $\mathcal{Q}$ is applied repeatedly to rotate the state vector within the subspace spanned by the good and bad components, gradually amplifying the probability of the target state. (3) QPE estimates the eigenvalue associated with this rotation by applying an inverse QFT to the evaluation register that has accumulated phase information through controlled applications of $\mathcal{Q}$. Measurement then yields a bitstring encoding an approximation of the rotation angle, from which the amplitude estimate $\tilde{a}$ is computed. (4) $\tilde{a}$ is classically rescaled to recover the expected path travel time. Due to the quadratic speed-up inherent in QAE, only $O(1/\epsilon)$ applications of the core quantum circuit are required to achieve $\epsilon$ with high probability, compared to $O(1/\epsilon^2)$ samples for classical Monte Carlo methods.

### *D. QAE-Based Path Travel Time Estimation*

1) **Quantum Representation of Path Travel Time**

First, we map the classical STN path estimation problem onto a quantum architecture. The total travel time state space is structured within a composite Hilbert space $\mathcal{H} = \mathcal{H}_\mathcal{J} \otimes \mathcal{H}_\mathcal{D} = \mathbb{C}^{k^n} \otimes \mathbb{C}^2$, comprising an index system $\mathcal{H}_\mathcal{J}$ and a data system $\mathcal{H}_\mathcal{D}$. The index system $\mathcal{H}_\mathcal{J}$ uses a $n$-qubit register to uniquely identify each joint realization of the path travel time. Each realization is identified by an index vector $j = (j_1, \ldots, j_n)^T \in K^n$, where each element corresponds to a discrete travel time of one link as defined in Section III.B. The corresponding basis state is formulated as the tensor product $|j\rangle = |j_n, \ldots, j_1\rangle = |j_n\rangle \otimes \cdots \otimes |j_1\rangle$, where $|j_l\rangle$ for $j_l = 0, \ldots, k-1$ is the orthogonal basis for the link $l$ [27]. The data system $\mathcal{H}_\mathcal{D}$, initialized in state $|\psi\rangle$, stores the accumulated path travel time associated with the realization indexed by $|j\rangle$.

The encoding process applies a realization-specific unitary operator $U(j): \mathbb{R}^n \to B(\mathcal{H}_\mathcal{J} \otimes \mathcal{H}_\mathcal{D})$ parameterized by the link travel time vector $x(j)$ to the data system. In the circuit, $U(j)$ is implemented through a controlled operator $c - U(j)$, which executes $U(j)$ only when the index register is in state $|j\rangle$ and acts as the identity operator otherwise. That is, $c - U(j) = |j\rangle\langle j| \otimes \left(V(x_{n,j_n}) \cdots V(x_{1,j_1})\right) + |j\rangle\langle j|_\perp \otimes I_\mathcal{D}$. Here, $V: \mathbb{R} \to \mathcal{H}_\mathcal{D}$ maps travel time values to the data system using controlled $R_y(\theta_l)$ rotation gates, where the per-link rotation angle $\theta_l$ is derived from the realized link travel time $t_l$, the accumulated angle $\theta_h$ corresponds to the path travel time of realization $j$. This mapping encodes the travel time information into the amplitude of the data qubit through rotation. The explicit construction of $c - U(j)$ is detailed in Section III.D.3).

The sequence of conditional operations yields the entangled state representing the STN:

$$|\Psi_f\rangle = \sum_{j \in K^n} p(j)|j\rangle \otimes U(j)\,|\psi\rangle, \tag{3}$$

where $p(j)$ is the amplitude associated with realization $j$, and its squared magnitude gives the joint probability $p^2(j) = \mathbb{P}[T_1 = x_{1,j_1}, \ldots, T_n = x_{n,j_n}] = P[T = x(j)]$. The expected path travel time is then recovered by evaluating an observable $M$ acting on the data system:

$$\langle M\rangle = \langle\Psi_f|\mathcal{H}_\mathcal{J} \otimes M|\Psi_f\rangle = \sum_{j \in K^n} p^2(j)\langle \mathrm{M(j)}\rangle_\psi \tag{4}$$

where: $\mathrm{M(j)} = U^{\dagger}(j)MU(j)$, and $\langle \mathrm{M(j)} \rangle = \sum_{l=1}^{n} x_{l,j_l}$ is the path travel time under realization $j$.

To clarify the encoding mechanism above, consider a simple path consisting of two links, (0,1) and (1,2) (i.e., $n = 2$). Each link has two discrete states (i.e., $k = 2$) of 10 and 20 minutes with a uniform probability of 0.5, as step-by-step parameterized below.

(1) Enumerations: The network yields $k^n = 2^2 = 4$ joint realizations of the path travel time. The results for each realization are summarized in Table II.

TABLE II

INFORMATION OF REALIZATIONS FOR PATH TRAVEL TIME.

| Realization ($\omega$) | Link Travel Time ($x(j)$) | | Path Travel Time ($T_h$/t($\omega$)) | Joint Probability ($P(\omega)$) | Amplitude $p(j)$ |
|---|---|---|---|---|---|
| | $t_{1,j_1}$ | $t_{2,j_2}$ | | | |
| 1 | 10 | 10 | $10 + 10 = 20$ | $0.5 * 0.5 = 0.25$ | 0.5 |
| 2 | 10 | 20 | $10 + 20 = 30$ | $0.5 * 0.5 = 0.25$ | 0.5 |
| 3 | 20 | 10 | $20 + 10 = 30$ | $0.5 * 0.5 = 0.25$ | 0.5 |
| 4 | 20 | 20 | $20 + 20 = 40$ | $0.5 * 0.5 = 0.25$ | 0.5 |

(2) Index Encoding: For $n = 2$, the index vector is $j = (j_1, j_2)^T$. Each realization $\omega$ is encoded by a 2-qubit basis state $|j\rangle = |j_2\rangle \otimes |j_1\rangle$. Specifically, the four realizations are mapped as follows: realization 1 corresponds to $|00\rangle$, realization 2 to $|01\rangle$, realization 3 to $|10\rangle$, and realization 4 to $|11\rangle$.

(3) Data Encoding: The controlled operator $c - U(j)$ applies an $R(\theta)$ rotation onto the data qubit initialized at the initial state $|0\rangle$, where the rotation angle directly correlates with the path travel time as summarized in Table III.

TABLE III

ROTATION OPERATION ON DATA SYSTEM OF REALIZATIONS.

| Realization ($\omega$) | Path Travel Time ($T_h$/t($\omega$)) | Rotation Operation on Data System $U(j)|\psi\rangle$ |
|---|---|---|
| 1 | $10 + 10 = 20$ | $R(\theta_{20})|0\rangle$ |
| 2 | $10 + 20 = 30$ | $R(\theta_{30})|0\rangle$ |
| 3 | $20 + 10 = 30$ | $R(\theta_{30})|0\rangle$ |
| 4 | $20 + 20 = 40$ | $R(\theta_{40})|0\rangle$ |

(4) Final State: Based on (3), the final superposition state can be expressed as: $|\Psi_{\exp}\rangle = 0.5|00\rangle \otimes R(\theta_{20})|0\rangle + 0.5|01\rangle \otimes R(\theta_{30})|0\rangle + 0.5|10\rangle \otimes R(\theta_{30})|0\rangle + 0.5|11\rangle \otimes R(\theta_{40})|0\rangle$. Here, the coefficient 0.5 matches the square root of the joint probability ($\sqrt{0.25} = 0.5$). QAE operates directly on this superposition to compute the amplitude, enabling the extraction of the expected path travel time $\mathbb{E}[T_h]$.

It is important to note that the formulation and example above assume independent link travel times. As such, the joint probability of realizations can be factorized into the product of link-level probabilities. This allows each link distribution to be encoded separately, and the index register to be constructed as a tensor product of independent subsystems, simplifying both index register and the controlled operations for encoding realization-dependent travel times.

The framework can also be extended to correlated link travel times by replacing independent distributions with a joint distribution. In this case, the index state must be prepared by a unified unitary operator that directly encodes all correlated joint realizations. The conditional operators would then act on joint basis states rather than separable link-level states. Although the overall framework remains applicable, state preparation and conditional encoding become more complex due to the higher-dimensional correlation structure, resulting in increased representation complexity.

2) **Measurement Scheme for Expected Path Travel Time**

Building on the encoding structure above, this subsection describes how the encoded information is converted into a measurable quantity that estimates $\mathbb{E}[T_h]$. Choosing the data-encoding operator $V(x)$ as $V(x) = R_y(\theta_l) = \cos(x/2)\, I - i\sin(x/2)\sigma_y$ and initializing the data system in $|0\rangle$ [27], the final state in (3) decomposes as:

$$|\Psi_{\mathrm{f}}\rangle = |\Psi_0\rangle + |\Psi_1\rangle \tag{5}$$

where $|\psi_0\rangle = \sum_{j \in K^n} p(j) \cos\left(\frac{1}{2} sum\{x(j)\}\right) |j\rangle|0\rangle$, $|\psi_1\rangle = \sum_{j \in K^n} p(j) \sin\left(\frac{1}{2} sum\{x(j)\}\right) |j\rangle|1\rangle$.

Here $sum\{x(j)\}$ denotes the realized path travel time under scenario $j$, and $|0\rangle$ and $|1\rangle$ components reflect how $sum\{x(j)\}$ controls the rotation applied to the data qubit. The measurement scheme therefore aggregates all path realizations into a single qubit whose amplitude carries their combined effect, so that the probability of measuring $|1\rangle$ encodes how the distribution of realizations shifts the data qubit away from its initial amplitude. Given the QAE estimate $\tilde{a}$, the expectation value is recovered as:

$$\mathbb{E}[\cos(T_h)] = 1 - 2\tilde{a} \tag{6}$$

3) **Quantum Circuit Design**

The superposition state in the previous subsection, together with the associated measurement scheme, provides a functional representation of the path travel time across all realizations. To implement this representation on a quantum device, the circuit must efficiently construct the index state and apply the controlled data rotations. Under the link-independence assumption, $\mathbb{E}[T_h]$ can be simplified to:

$$\mathbb{E}[T_h] = \sum_{j \in K^n} sum\{x(j)\} \prod_{l=1}^{n} P[X_l = x_{l,j_l}]. \tag{7}$$

This structure partitions the index system $\mathcal{H}_J$ into $n$ independent level-index subsystems, $\mathcal{H}_J = \otimes_{l=1}^{n} \mathcal{H}_{Jl}$, where each subsystem $\mathcal{H}_{Jl} = \mathbb{C}^k$ corresponds to the $k$ realizations of link $l$. The partition also shows the similar structure of the classical model, in which each link contributes its own random travel time and the path-level outcome is obtained by their combination. The global index state $|\varphi\rangle$ is thus prepared by a tensor product of independent link states: $|\varphi\rangle = \sum_j p(j)|j\rangle = \otimes_{l=1}^{n} \left(\sum_{j_l=0}^{k-1} p_l(j_l)|j_l\rangle\right)$.

The state $|\varphi\rangle$ is prepared by a sequence of unitary operators

$A = A_n \dots A_1$, in which each $A_l$ acts only on its corresponding subsystem $\mathcal{H}_{\mathcal{J}l}$ to load the probability distribution for link $l$. Since these operators act on separate subsystems, they commute and can be applied in parallel. Once $|\varphi\rangle$ is prepared, it is entangled with the data system via the total unitary operator $U(j) = U_n(x_n) \cdots U_1(x_1)$, where each $U_l(x)$ executes sequential multi-controlled rotations conditioned on the corresponding index register state: $U_l(x) = \prod_{j=0}^{k-1} c - V_{lj_l}(x_j) \in B(\mathcal{H}_{\mathcal{J}} \otimes \mathcal{H}_{\mathcal{D}})$. The controlled rotation operator $c - V_{lj_l}(x_j)$ is mathematically formulated through a localized projection operator:

$$c - V_{lj_l}(x) \coloneqq \prod_{lj_l} \otimes V(x) + \prod_{lj}^{\perp} \otimes I, \tag{8}$$

where $\prod_{lj} = |j\rangle\langle j|_l \triangleq I_k^{\otimes l-1} \otimes |j\rangle\langle j| \otimes I_k^{\otimes n-l}$ projects onto the target state $|j\rangle$ of the $l$-th link subsystem.

These operators ensure that when the link- $l$ index subsystem is in state $|j_l\rangle$, the controlled operator selects the associated travel time value and applies the corresponding rotation to the data register. The circuit thus preserves the link-level decomposition of the distribution while converting the combined effect of all link travel times into the quantum amplitude of the data system, enabling the measurement scheme to extract the desired statistical quantity from the superposition. Fig. 2 provides the block-level schematic of this circuit structure: the index system is prepared $A = A_n \dots A_1$, with each $A_l$ acting on $\mathcal{H}_{\mathcal{J}l}$, and the sequential link-level travel time accumulation is implemented by multi-controlled blocks $U(j) = U_n(x_n) \cdots U_1(x_1)$, with each $U_l(x_l)$ controlled by $\mathcal{H}_{\mathcal{J}l}$ and acting on the data register $\mathcal{H}_{\mathcal{D}}$.

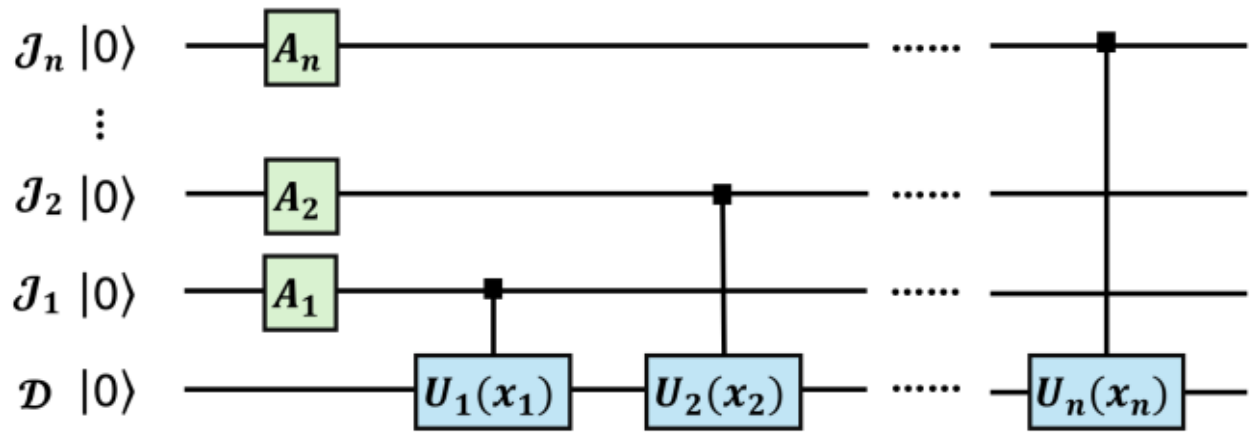

**Fig. 2.** Circuit design for QAE-based path expected travel time estimation.

*E. QAE Variants and Performance Trade-offs*

Several QAE variants have been developed with different trade-offs in circuit depth, ancilla requirements, estimation precision, and implementation complexity. We focus on four prominent approaches: Canonical AE (CAE), iterative AE (IAE), maximum likelihood AE (MLAE), and faster AE (FAE), comparing their mechanisms, resource requirements, and performance characteristics as summarized in Table IV.

CAE combines QPE [34] with amplitude amplification, applying controlled powers of the Grover operator followed by an inverse QFT [35] to estimate the target amplitude [11]. It requires multiple ancilla qubits and deep circuits, and therefore offers strong theoretical performance but is difficult to implement on quantum hardware with limited coherence.

TABLE IV
SUMMARY OF QAE VARIANTS.

| Method | Based on QPE | Ancilla count | Strength | Limitation |
|---|---|---|---|---|
| CAE | Yes | High | High accuracy | Deep circuits |
| IAE | No | Low | Noise-resistant | Iterative |
| MLAE | No | Low | Statistically efficient | Complex likelihood fitting |
| FAE | Reduced | Low | Fast runtime | Lower precision |

IAE replaces QPE with an adaptive procedure that performs a sequence of measurements at selected amplification levels [14]. By iteratively updating a confidence interval for the target amplitude, IAE decreases ancilla requirements and circuit depth, providing enhanced noise resilience on near-term quantum devices. MLAE eliminates QPE by executing a series of amplitude- amplification circuits with varying iteration numbers and using the resulting measurement statistics to form a likelihood function [15]. It optimizes data efficiency and maintains hardware-compatible circuits, though its accuracy depends heavily on the optimization stability of the likelihood function. Finally, FAE applies amplitude amplification across strategically selected schedules, resolving cosine phase ambiguity by merging multi-frequency measurements through an iterative refinement rule [16]. It has a reduced runtime with short-depth circuits and minimal ancilla overhead, but loses precision relative to CAE.

*F. Rotation-Angle Scaling and Post-processing*

As discussed in Sections III.D.1) and III.D.3), the data-encoding operator $V(x)$ is implemented as $R_y(\theta)$ rotation gate which maps a physical travel time value to a bounded rotation angle. This rotation determines the state register's projection following the application of $U(j)$, directly controlling how different realizations contribute to the amplitude measured by QAE. Since the mapping from physical time to the rotation domain affects the curvature of the cosine function in (5), inappropriate scaling may distort the differences among realizations and degrade estimation precision. Rotation-angle scaling is therefore introduced to map travel time values into a numerically stable region of the trigonometric interval, ensuring that the QAE output can be consistently post-processed to recover $\mathbb{E}[T_h]$.

We employ and evaluate four scaling methods, each paired with a specific post-processing procedure. Scaling Method 1 directly uses the raw link travel time realization $t_l$ without normalization. Scaling Methods 2-4 normalize the rotation angle into $[0, \pi]$ using either path-level or link-level range information.

Scaling Method 1 directly sets $\theta_l = t_l$ without scaling, preserving the initial magnitude data within the $R_y(\theta_l)$ operation. After obtaining $\mathbb{E}[\cos(T_h)]$ from (6), post-processing computes the estimated angle $\widetilde{\theta_h}$ through $\widetilde{\theta_h} = \arccos(\mathbb{E}[\cos(T_h)])$. A correction term $c = \mathbb{E}[\theta_h] - \widetilde{\theta_h}$, with $\mathbb{E}[\theta_h] = \sum_{j \in K^n} p^2(j)\, \theta_j$, is then added to obtain the corrected angle $\theta_h^* = \widetilde{\theta_h} + c$, and finally $\mathbb{E}[T_h] = \theta_h^*$. Scaling Method 2 uses a path-level $\gamma = \pi/(S_{max}^h - S_{min}^h)$, where $S_{max}^h$ and $S_{min}^h$ are the maximum and minimum total travel times of path $h$. The rotation angle is $\theta_l = \left(t_l - S_{min}^h/n\right) \cdot \gamma$, and the post-processing formula is $\mathbb{E}[T_h] = \theta_h^*/\gamma + S_{min}^h$. Scaling Method 3 shares $\gamma = \pi/(S_{max}^h - S_{min}^h)$ but normalizes the rotation angle using the link-specific minimum travel time $t_{min}^l$ as $\theta_l = \left(t_l - t_{min}^l\right) \cdot \gamma$. This focuses on scaling at the link level and is suitable when link-level variations dominate the variability of the total path. The post-processing formula is $\mathbb{E}[T_h] = \theta_h^*/\gamma + T_{min}^h$, where $T_{min}^h$ is the sum of the minimum travel times of all links along the path. Scaling Method 4 applies a link-level scaling factor $\gamma = \pi/(t_{max}^l - t_{min}^l)$, where $t_{max}^l$ and $t_{min}^l$ are the maximum and minimum travel times of link $l$, with $\theta_l = \left(t_l - t_{min}^l\right) \cdot \gamma$. This preserves link-level variability and is appropriate when links have substantially different distributions. The post-processing formula is $\mathbb{E}[T_h] = \sum_{l=1}^{n} \frac{\theta_h^*}{n}/\gamma + t_{min}^l$.

### *G. Discussion of Computational Scalability and Limitations*

Although QAE offers a theoretical quadratic improvement in convergence rate, its practical performance in stochastic routing is currently constrained by the limited qubit capacity, restricted circuit depth, and noise characteristics of NISQ hardware. These limitations directly affect the reliability of state preparation and amplitude encoding. This subsection therefore discusses the computational scalability and practical feasibility of the proposed QAE-based estimator.

For the classical Monte Carlo introduced in Section III.A, when the output variance is bounded by $\sigma^2$, the empirical estimator $\tilde{\mu}$ deviates from the true mean $\mu$ by more than error $\varepsilon$ by Chebyshev's inequality: $\Pr[|\tilde{\mu} - \mu| \geq \varepsilon] \leq \frac{\sigma^2}{N\varepsilon^2}$. This requires $N = O(\sigma^2/\varepsilon^2)$ evaluations to achieve $\varepsilon$ with constant success probability [24]. For $\sigma = 1$ and $\varepsilon = 10^{-3}$, this corresponds to roughly $10^6$ executions, imposing a substantial cost for repeated expectation evaluation.

By contrast, QAE achieves $\varepsilon$-accuracy with only $O(\sigma/\varepsilon)$ quantum queries [12]. Since $\sigma$ typically does not grow with problem size in practical STN settings, this query complexity simplifies to $O(1/\varepsilon)$, yielding an asymptotic comparison of $O(1/\varepsilon^2)$ for Monte Carlo versus $O(1/\varepsilon)$ for QAE. This theoretical quadratic speed-up is well-established in stochastic quantum simulations [12], [24], [27]. However, we emphasize that such theoretical speedups are used in this study to motivate the algorithmic framework rather than to claim near-term computational superiority on current hardware. Realizing the advantage at scale remains contingent on advances in qubit counts, coherence times, and gate error rates.

Beyond asymptotic complexity, practical scalability depends primarily on the qubit count and circuit depth required for state preparation and amplitude encoding. For a path with $n$ links and $k$ discrete travel time values per link, the index register from Section III.D.1) requires approximately $n_{index} \approx n log_2 k$ qubits. Together with the data qubit and ancilla qubits, the total qubit count is $Q = n_{index} + n_{data} + n_{ancilla}$, where the data register encodes travel time contributions and the ancilla register supports amplitude encoding and controlled reflection. Consequently, the qubit requirement scales linearly with path size and logarithmically with the discretization resolution.

Current NISQ hardware [9] typically supports shallow circuits with fewer than 20 reliably controllable qubits, which places a practical upper bound on the implementable path size and discretization level. Furthermore, quantum circuit depth is limited by the cumulative noise of the controlled rotations in travel time encoding and iterative structure of QAE. While the unitary state preparation scales linearly, the two-qubit gates required for conditional operations introduce significantly higher error rates than single-qubit operations, representing a primary source of hardware infidelity [36]. Given these hardware constraints, practical deployment will require hybrid quantum-classical architectures in which QAE is selectively applied to computationally intensive expectation-evaluation subprocesses while classical algorithms handle global routing and feasibility checking. Within this division of computational roles, the proposed framework supplies QAE-based travel time estimates to classical routing solvers, reducing reliance on large-scale sampling while preserving compatibility with existing ITS optimization workflows.

Ultimately, this study provides a proof-of-concept validation of applying QAE to stochastic routing and demonstrates its potential for ITS applications. While the theoretical quadratic speed-up is mathematically verified, the practical realization on large-scale networks remains restricted by the physical capabilities of NISQ devices. Consequently, the immediate value of this work lies in establishing the algorithmic foundations and travel time encoding strategies. As quantum hardware matures to support higher qubit counts and error correction, the proposed QAE-based estimator will be ready to address complex and large-scale stochastic routing problems that are currently intractable for classical computing.

## IV. Experimental Setup and Design

This section presents the experimental setup used to evaluate the proposed QAE-based estimator. It first introduces a small-scale STN case study and then specifies the simulation environment together with two experimental designs.

### *A. STN Case Study*

We construct a small-scale STN consisting of a central depot (node 0) and three customer nodes (nodes 1, 2, and 3). The network has independent discrete random travel times across all directed links. Three parametric distributions are used to generate link travel times: uniform, normal, and lognormal. For each link, three possible travel time values are drawn within the range of 20 to 50 minutes. The probabilities for the uniform distribution are set identically at $1/3$, while the probabilities for the normal and lognormal distributions are derived from their cumulative distribution functions across the discretized intervals.

### *B. Experimental Design*

All experiments are conducted on a local IBM Qiskit simulator, with each quantum circuit executed by the Qiskit sampler [37] using the default setting of 1024 shots. The four QAE variants are configured with parameters that balance estimation precision and simulator cost: CAE uses 5 evaluation qubits. IAE uses a target precision of 0.1 and a confidence level of 0.05. MLAE adopts the evaluation schedule [0, 1, 2, 4], specifying the powers of the Grover operator applied at successive stages. FAE sets the failure probability to 0.1 with at most 5 iterations.

Two experiments are conducted on this setup. **Experiment 1** compares the accuracy of CAE, IAE, MLAE, and FAE on all candidate paths under the three travel time distributions, using the four scaling methods defined above at the baseline rotation-angle range $[0, \pi]$. **Experiment 2** assesses the sensitivity of estimation accuracy to the choice of rotation-angle range. For each path, distribution, and QAE variant, each scaling method is combined with five rotation-angle ranges growing from $[0, \pi]$ to $[0, 5\pi]$ in increments of $\pi$.

## V. Results and Analysis

This section presents the experimental results in two parts. Section V.A reports Experiment 1, and Section V.B reports Experiment 2. The discussion focuses on overall patterns and robustness rather than absolute errors for individual paths.

### *A. Results of the QAE Method Performance Comparison*

Experiment 1 evaluates the absolute error between the QAE-based estimates and the ground-truth expected path travel times. Fig. 3 presents these per-path errors the three distributions and four scaling methods. To support the discussion, we also compute the mean, median, standard deviation, minimum, and maximum of the absolute errors across the six paths are computed for each (distribution, scaling method, QAE variant) combination. Across all configurations, the four QAE variants produce comparable per-path errors, ranging from 0.0058 to 1.0531.

Scaling Method 1 consistently yields the largest errors across the three distributions, indicating that all three scaling strategies substantially outperform no scaling. At the path level, IAE attains the smallest absolute error most frequently (25 of 72 configurations, counting ties), followed by CAE (24), MLAE (18), and FAE (8), but no single variant is uniformly superior.

Distribution-specific behavior is also visible in Fig. 3. Under the uniform distribution, CAE attains the smallest mean error under Scaling Methods 1 and 2 (0.699 and 0.358), while the four variants become nearly equivalent under Methods 3 and 4. Under the normal distribution, IAE retains a small but consistent advantage across all four scaling methods, with a mean of 0.585 under Scaling Method 1.

The lognormal distribution shows the largest cross-path variability under Scaling Method 1, where MLAE and FAE achieve the lowest means (0.485 and 0.484), while CAE exhibits the strongest path dependence. Under Methods 2-4,

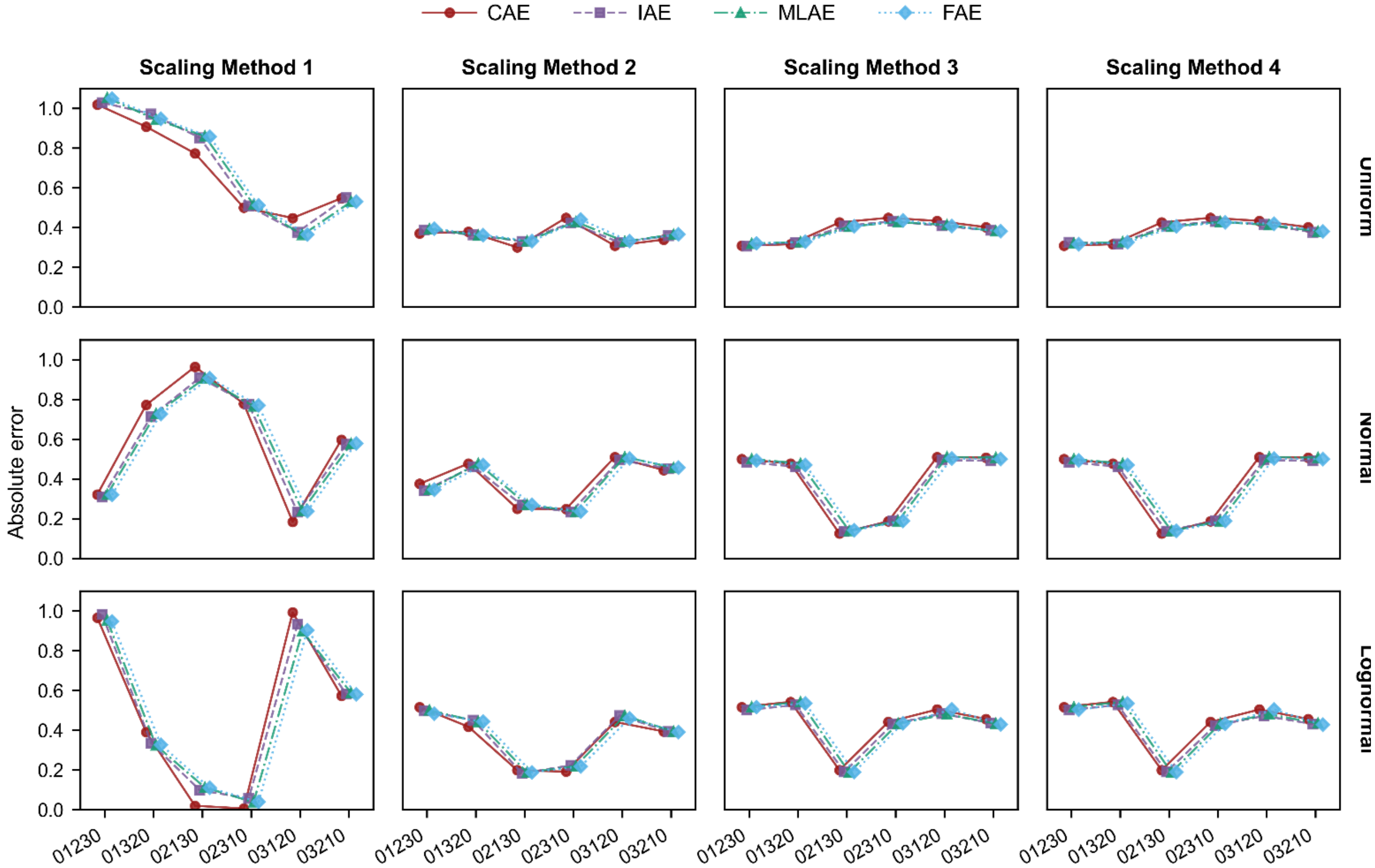

**Fig. 3.** Absolute errors of the four QAE methods at rotation-angle range $[0, \pi]$ under three travel time distributions.

the four variants again converge within 0.02.

Overall, QAE provides consistently accurate path-level travel time estimates across the three distributions, supporting its use as an alternative to Monte Carlo simulation for expectation estimation in small-scale STN settings. IAE offers the most stable behavior overall, but the dominant determinant of accuracy is the scaling design rather than the choice of QAE variant itself.

### *B. Results of the Rotation Angle Scaling Range Impact Analysis*

Experiment 2 evaluates the sensitivity of rotation-angle range by computing the mean absolute error of each QAE method across all paths under each combination of scaling method and rotation angle range, for all three distributions. Fig. 4 presents these results via heatmaps, with rows showing scaling method and rotation-angle range pairs and columns showing the four QAE variants; panels (a)-(c) correspond to the uniform, normal, and lognormal distributions.

The mean absolute errors for all scaling ranges remain relatively small, ranging from 0 to 0.45. The rotation-angle range has a dominant influence on estimation accuracy, with a consistent pattern across the three distributions. Ranges corresponding to even multiples of $\pi$, namely $[0, 2\pi]$ and $[0, 4\pi]$, yield substantially lower mean errors than ranges corresponding to odd multiples $[0, \pi]$, $[0, 3\pi]$, and $[0, 5\pi]$. The darkest cells of Fig. 4 are concentrated in the $[0, \pi]$ rows and the no-scaling row of Scaling Method 1, while the lightest cells appear at the intersection of $[0, 2\pi]$ or $[0, 4\pi]$ with Scaling Methods 3 and 4, where mean errors approach zero for CAE and MLAE and remain below 0.008 for IAE and FAE. Scaling Method 2 produces low errors at the same even multiples (0.010 to 0.053 across distributions) but consistently higher than those of Scaling Methods 3 and 4. Methods 3 and 4 yield almost identical heatmap patterns across all rotation-angle ranges, while differences among the four QAE variants within any given cell remain within 0.01.

In summary, Experiment 2 shows that estimation accuracy is mainly affected by the scaling method and rotation-angle range. Pairing Scaling Method 3 or 4 with $[0, 2\pi]$ or $[0, 4\pi]$ performs best, with CAE and MLAE achieving the lowest errors. These findings confirm that the interaction between rotation-angle and scaling method plays a more critical role in estimation accuracy than the choice of QAE variant under the tested conditions.

## VI. Conclusion

This study proposes and validates a QAE-based estimator for expected path travel times in STNs, presenting the first application of QAE to a core bottleneck in stochastic routing: the repeated evaluation of stochastic travel times. By replacing classical Monte Carlo-based simulation with a quantum estimation procedure, the framework offers three main advantages. First, it provides a theoretical quadratic speed-up, reducing the complexity of expectation evaluation from $O(1/\varepsilon^2)$ under classical Monte Carlo sampling to $O(1/\varepsilon)$. Second, it requires no prior assumptions on the underlying travel time distributions, enhancing its generality in real-world network scenarios where distribution parameters are often unknown. Third, it encodes all possible travel time realizations into a quantum superposition, reducing the reliance on large-scale random sampling that dominates the cost of Monte Carlo-based methods.

Small-scale numerical experiments confirm the feasibility of the proposed approach, with all four QAE variants achieving consistently low estimation errors. A key finding is that the estimation precision depends primarily on the interaction between scaling method and rotation-angle range rather than on the choice of QAE variant alone. Among the tested variants, IAE exhibits the highest stability, while CAE remains competitive in several cases. Specifically, the combination of Scaling Methods 3 or 4 with rotation-angle ranges $[0, 2\pi]$ or $[0, 4\pi]$ consistently yields the lowest errors, providing configuration guidance for ITS applications.

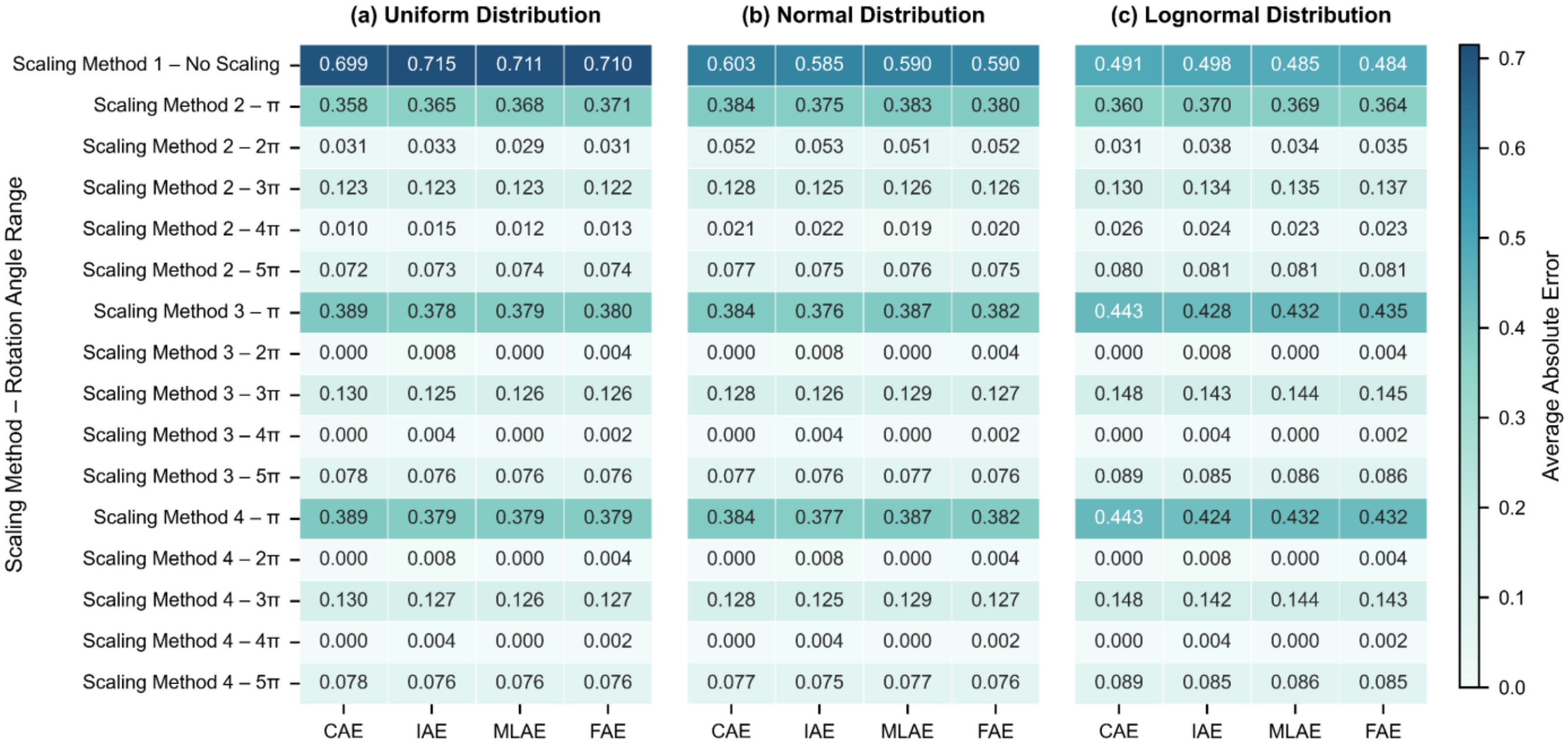

**Fig. 4.** Heatmaps of mean absolute error under uniform, normal, and lognormal travel time distributions.

These findings suggest that QAE can serve as an expectation-estimation module within simulation-based transportation models, particularly when repeated sampling dominates the computational cost. From an ITS perspective, this capability is relevant to stochastic routing, dynamic logistics planning, and real-time decision support systems in which repeated travel time evaluation can become a computational bottleneck.

Future work should investigate the integration of the proposed estimator with classical optimization solvers. In such a hybrid quantum-classical structure, QAE could be used to rapidly evaluate candidate path costs and may substantially improve the efficiency of solving the VRPSTT. This structure is consistent with practical ITS deployment, where quantum computing is more likely to support specific computation-intensive modules than to replace existing traffic management and logistics optimization systems.

Finally, the promising results obtained here are limited by the capabilities of current NISQ devices. Restricted qubit counts limit the path size and discretization resolution that can be supported, and quantum noise degrades circuit fidelity and estimation accuracy. Future research should therefore evaluate the proposed framework on larger and more realistic networks, incorporate correlated and time-dependent travel time distributions, and test implementation on noisy quantum hardware. Such work will be essential for determining the potential role of QAE in next generation ITS.